\documentclass[12pt,epsf]{article}

\newcommand{\dlt}{\delta}

\newcommand{\gm}{\gamma}
\newcommand{\Gm}{\Gamma}
\newcommand{\tht}{\theta}
\newcommand{\btht}{\bar{\tht}}
\newcommand{\kp}{\kappa}
\newcommand{\lmd}{\lambda}
\newcommand{\Lmd}{\Lambda}
\newcommand{\sgm}{\sigma}
\newcommand{\vph}{\varphi}

\newcommand{\be}{\begin{equation}}
\newcommand{\ee}{\end{equation}}
\newcommand{\bea}{\begin{eqnarray}}
\newcommand{\eea}{\end{eqnarray}}
\newcommand{\eql}{\!\!\!&=\!\!\!&}

\newcommand{\defa}{\!\!\!&\equiv\!\!\!&}

\newcommand{\tl}[1]{\tilde{#1}}

\newcommand{\diag}{{\rm diag}}
\newcommand{\der}{\partial}
\newcommand{\dr}{\!\!{\rm d}}

\newcommand{\Acl}{A_{\rm cl}}

\newcommand{\brkt}[1]{\left( #1 \right)}
\newcommand{\brc}[1]{\left\{ #1 \right\}}
\newcommand{\sbk}[1]{\left[ #1 \right]}

\renewcommand{\Re}{{\rm Re}}
\renewcommand{\Im}{{\rm Im}}

\newcommand{\cD}{{\cal D}}
\newcommand{\cF}{{\cal F}}

\newcommand{\cL}{{\cal L}}

\newcommand{\cN}{{\cal N}}
\newcommand{\cO}{{\cal O}}

\newcommand{\NP}[1]{{\it Nucl.~Phys.}~{\bf #1}}
\newcommand{\PL}[1]{{\it Phys.~Lett.}~{\bf #1}}

\newcommand{\PR}[1]{{\it Phys.~Rev.}~{\bf #1}}
\newcommand{\PRL}[1]{{\it Phys.~Rev.~Lett.}~{\bf #1}}
\newcommand{\PTP}[1]{{\it Prog.~Theor.~Phys.}~{\bf #1}}

\newcommand{\JH}[1]{{\it JHEP}~{\bf #1}}

\begin{document}

\begin{titlepage}
\null
\begin{flushright}
 {\tt hep-th/0302196}\\
TU-681
\\
Feb 2003
\end{flushright}

\vskip 2cm
\begin{center}
{\LARGE \bf  Modified mode-expansion on a BPS wall \\
related to the nonlinear realization}

\lineskip .75em
\vskip 2.5cm

\normalsize

{\large \bf Yutaka Sakamura}
{\def\thefootnote{\fnsymbol{footnote}}
\footnote[5]{\it  e-mail address:
sakamura@tuhep.phys.tohoku.ac.jp}}

\vskip 1.5em

{\it Department of Physics, Tohoku University\\ 
Sendai 980-8578, Japan}

\vspace{18mm}

{\bf Abstract}\\[5mm]
{\parbox{13cm}{\hspace{5mm} \small
We propose a modified mode-expansion of the bulk fields in a BPS domain wall 
background to obtain the effective theory on the wall. 
The broken SUSY is nonlinearly realized on each mode 
defined by our mode-expansion. 
Our work clarifies a relation between two different approaches 
to derive the effective theory on a BPS wall, 
{\it i.e.} the nonlinear realization approach 
and the mode-expansion approach.  
We also discuss a further modification that respects the Lorentz 
and $U(1)_R$ symmetries broken by the wall. 
}}

\end{center}

\end{titlepage}

\clearpage

\section{Introduction}
In supersymmetric (SUSY) theories, there are important states called 
BPS states \cite{BPS}. 
They preserve part of the supersymmetry of the theory 
and play a crucial role in quantum field theories. 
One of the simplest example of BPS states is a BPS domain wall. 
In particular, BPS domain walls in four-dimensional (4D) $\cN=1$ SUSY 
theories have been thoroughly investigated in a number of papers 
because such theories are tractable and have various types of 
BPS walls with interesting features 
\cite{Dvali,Chibisov,Gabadadze,Hou,Naganuma}. 

Besides that BPS domain walls are an intriguing subject to research 
in their own rights, they are also important 
in the brane-world scenario \cite{Rubakov,Arkani,Randall} 
because they can provide a natural realization 
of 5D $\cN=1$ SUSY (eight supercharges) to 4D $\cN=1$ SUSY 
(four supercharges), which is relevant to the phenomenology. 
Of course, since our world is four-dimensional, we should discuss 
a domain wall in 5D theories for the realistic model-building. 
However, 5D SUSY theory is quite restrictive and difficult to handle. 
So it is useful and instructive to study BPS walls 
in 4D $\cN=1$ theories as a toy model. 
In this paper, we will concentrate ourselves on BPS walls 
in the 4D $\cN=1$ generalized Wess-Zumino model for simplicity. 
To discuss the physics in the BPS-wall background,  
it is useful to investigate a low-energy effective theory on the wall. 
Since a BPS wall preserves a half of the bulk SUSY, 
such an effective theory can be expressed by 3D superfields. 
There are mainly two approaches to derive the effective theory 
described by 3D superfields. 

The first one is the nonlinear realization approach \cite{ivanov1,bagger1}. 
From the 3D viewpoint, SUSY breaking by a BPS wall can be regarded as 
the partial SUSY breaking from 3D $\cN=2$ to $\cN=1$. 
Thus, we can obtain the effective theory on the wall 
by constructing an invariant action under the broken SUSY, 
which is realized nonlinearly. 
This approach is useful for the discussion of the general properties 
of BPS supermembranes 
since this method uses only information on the breaking pattern 
of the symmetries. 
This also means, however, that we cannot determine parameters 
of the effective theory in this approach. 
 
The second one is the mode-expansion approach \cite{sakamura1}. 
In this approach, the 3D effective theory is directly derived from 
the original 4D theory. 
Specifically, we expand the fluctuation fields around the wall-background 
into an infinite number of 3D superfields, and integrating out 
the heavy modes. 
In this approach, we can see explicitly how the 3D superfields 
in the effective theory are embedded into the original 4D superfields. 
Unlike the previous approach,  parameters of the effective theory are 
obtained as the overlap integrals 
of the background field configuration and the mode functions. 
However, this approach does not respect the symmetries broken by the wall. 

These two approaches are complementary to each other. 
So it is useful and instructive 
to clarify the relation between them. 
This is the purpose of this paper. 
Specifically, we will propose the modification of 
the naive mode-expansion of the bulk superfields 
so that the broken SUSY is nonlinearly realized on each mode. 
Using our modified mode-expansion, we can obtain an invariant 
effective action under the broken SUSY, and 
can also calculate parameters of the effective theory. 
In the latter part of this paper, we try to modify the mode-expansion 
further so that it also respects the Lorentz and $U(1)_R$ symmetries 
broken by the wall. 

The paper is organized as follows. 
In the next section, we will provide a brief review of 
the nonlinear realization approach. 
In Sec.\ref{MEA}, we will review our previous work \cite{sakamura1} 
where a naive mode-expansion is discussed. 
Then, in Sec.\ref{mdfd_MEA}, we will modify it so that the broken SUSY 
is nonlinearly realized on each mode. 
In Sec.\ref{bk_Lorentz}, we will discuss the further modification 
of the mode-expansion which also respects the broken Lorentz  
and $U(1)_R$ symmetries. 
Sec.\ref{summary} is devoted to the summary and the discussion. 
Notations and some formulae are listed in the appendices.

\section{Review of the nonlinear realization approach}
In this section, we will briefly review the nonlinear realization approach 
to construct an effective action for the supermembrane~\cite{ivanov1}. 
Throughout this paper, we will assume that the background space-time is 
flat and has a 4D $\cN=1$ supersymmetry. 

From the 3D viewpoint, the 4D $\cN=1$ SUSY algebra is 
a central extended $\cN=2$ SUSY algebra. 
\bea
 \{Q_{1\alpha},Q_{1\beta}\} \eql \{Q_{2\alpha},Q_{2\beta}\}
 =2(\gm_{(3)}^m\sgm^2)_{\alpha\beta}P_m, \nonumber\\
 \{Q_{1\alpha},Q_{2\beta}\} \eql -\{Q_{2\alpha},Q_{1\beta}\}
 =2i(\sgm^2)_{\alpha\beta}P_2, \label{SUSYalg1}
\eea
where $m=0,1,3$ denotes the 3D space-time index, and $\alpha,\beta$ denote
the 3D Majorana spinor indices\footnote{
In this paper, we will choose the $x_2$-direction to be perpendicular 
to the membrane or the domain wall.
}. 

The existence of the BPS membrane breaks the bulk 
symmetry~$G=\{P_m,P_2,Q_{1\alpha},Q_{2\alpha}\}$ to 
the vacuum stability subgroup~$H=\{P_m,Q_{2\alpha}\}$. 
Then, a coset element~$\Omega$ can be parameterized as 
\be
 \Omega = e^{ix^mP_m+\tht_2Q_2}e^{i\kp\rho_0P_2+\kp\zeta_0Q_1}. 
 \label{Omega}
\ee
Here $\rho_0=\rho_0(x^m,\tht_2)$ and 
$\zeta_{0\alpha}=\zeta_{0\alpha}(x^m,\tht_2)$ are 
scalar and spinor superfields corresponding to the Nambu-Goldstone 
(NG) modes for $P_2$ and $Q_{1\alpha}$ respectively, 
and $\kp$ is a constant\footnote{
The mass scale~$f\equiv \kp^{-2/3}$ corresponds to the scale 
where $P_2$ and $Q_1$ are broken, 
and $f^3=\kp^{-2}$ is a tension of the membrane.} 
whose mass-dimension is $-3/2$. 

The transformation laws of each superfield for the broken symmetries 
can be read off as follows by multiplying $\Omega$ by corresponding 
group elements from the left. 
\bea
 \dlt^{P_2}_a \rho_0 \eql a,  \nonumber\\
 \dlt^{P_2}_a \zeta_{0\alpha} \eql \dlt^{P_2}_a \phi =0, \label{dltZ}
\eea
\bea
 \dlt^{Q_1}_{\xi_1}\rho_0 \eql -2\kp^{-1}\xi_1\tht_2
  -i\kp\xi_1\gm_{(3)}^m\zeta_0\der_m\rho_0,  \nonumber\\
 \dlt^{Q_1}_{\xi_1}\zeta_{0\alpha} \eql \kp^{-1}\xi_{1\alpha}
  -i\kp\xi_1\gm_{(3)}^m\zeta_0\der_m\zeta_{0\alpha},  \nonumber\\ 
 \dlt^{Q_1}_{\xi_1}\phi \eql 
  -i\kp\xi_1\gm_{(3)}^m\zeta_0\der_m\phi,  \label{dltQ1}
\eea
where $a$ and $\xi_{1\alpha}$ are transformation parameters, 
and $\phi$ denotes a matter superfield. 
Indeed, the above transformations satisfy 
the SUSY algebra~(\ref{SUSYalg1}). 
In this paper, we will call the above transformation laws 
for the broken symmetries 
the {\it standard non-linear transformations}. 

Note that the NG superfields~$\rho_0$ and $\zeta_0$ introduced by 
Eq.(\ref{Omega}) are not independent of each other, 
because the NG modes for $P_2$ and $Q_1$ form a supermultiplet 
for the unbroken $Q_2$-SUSY. 
The relation between them can be obtained by setting a covariant constraint 
(inverse Higgs effect~\cite{ivsHiggs}), 
\be
 \cD_{2\alpha}\rho_0=0,  \label{cov-cstrt1}
\ee
where $\cD_{2\alpha}$ is a covariant spinor derivative in the presence of 
the NG superfields, and its explicit form is listed in Eq.(\ref{cov_drvs1}) 
in Appendix~\ref{cov_ders}. 
From this constraint, we can express $\zeta_{0\alpha}$ in terms of $\rho_0$. 
\be
 \zeta_{0\alpha}=-\frac{1}{2}D_{2\alpha}\rho_0+\cO(\kp^2).  \label{zeta-rho}
\ee
Namely, the essential NG superfield is $\rho_0(x^m,\tht_2)$ only.  

By using $\rho_0$, we can construct a 3D $\cN=2$ invariant action 
perturbatively for $\kp$ by the standard procedure of 
the nonlinear realization~\cite{bagger1}. 

For an invariant action for the NG modes~$S_{\rm NG}$, 
there is an alternative method to derive the effective action 
for all orders in $\kp$~\cite{ivanov1,bagger2}. 
We will explain this method in the following. 

Let us introduce a scalar NG superfield~$\vph_0(x^m,\tht_2)$. 
As will be shown, $\vph_0$ coincides with $\rho_0$  
at the lowest order in $\kp$. 
Then, define a spinor superfield~$\psi_{0\alpha}(x^m,\tht_2)$ as 
\be
 \psi_{0\alpha}\equiv -\frac{1}{2}D_{2\alpha}\vph_0.  \label{def_psi}
\ee
This means that $\psi_0$ satisfies the following constraint. 
\be
 D_2^2\psi_{0\alpha}=-2i(\gm_{(3)}^m\der_m\psi_0)_\alpha. 
 \label{psi_cnstrt}
\ee

Noticing that $\psi_{0\alpha}$ is the NG superfield for $Q_1$, 
we can write its $Q_1$-transformation law which preserves 
the constraint~(\ref{psi_cnstrt}) and satisfies 
the SUSY algebra~(\ref{SUSYalg1}) as 
\be 
 \dlt_{\xi_1}^{Q_1}\psi_{0\alpha}=\kp^{-1}\xi_{1\alpha}
 +\frac{\kp}{4}\xi_{1\alpha}D_2^2\Xi-\frac{i\kp}{2}
 (\gm_{(3)}^m\xi_1)_\alpha\der_m\Xi.  \label{dlt_psi}
\ee
Here $\Xi$ is a scalar superfield and transforms under 
the $Q_1$ transformation as 
\be
 \dlt_{\xi_1}^{Q_1}\Xi=2\kp^{-1}\xi_1\psi_0=-\kp^{-1}\xi_1D_2\vph_0. 
 \label{dlt_Xi}
\ee
From Eqs.(\ref{dlt_psi}) and (\ref{dlt_Xi}), we can see that $\Xi$ 
satisfies the following recursive equation. 
\be
 \Xi=\frac{\psi_0^2}{1+\frac{\kp^2}{4}D_2^2\Xi}. 
\ee
This equation can be solved and the solution is 
\be
 \Xi=\frac{2\psi_0^2}{1+\sqrt{1+\kp^2D_2^2(\psi_0^2)}}. 
 \label{expr_Xi}
\ee

From the transformation law~(\ref{dlt_Xi}), we can see that 
\be 
 S_{\rm NG}=\int\dr^3x{\rm d}^2\tht_2 \: 2\Xi  \label{Sng}
\ee
is $Q_1$-invariant. 
So this is a good candidate for the effective action of the NG modes. 
The right-hand-side of Eq.(\ref{Sng}) is expanded by $\kp$ as
\be
 S_{\rm NG}=\int\dr^3x{\rm d}^2\tht_2 \: 
 \left\{\frac{1}{2}(D_2\vph_0)^2+\cO(\kp^2)\right\}.
\ee
This certainly is an action for a massless superfield. 
Actually, after eliminating the auxiliary field of $\vph_0$, 
the bosonic part of $S_{\rm NG}$ becomes the Nambu-Goto action 
in the static gauge. (See Appendix~\ref{SUSY_NG}.) 

Now we define the quantity 
\be
 E\equiv \frac{1}{1+\frac{\kp^2}{4}D_2^2\Xi}. \label{def_E}
\ee
Then the relation between $\psi_{0\alpha}$ and $\zeta_{0\alpha}$ 
can be written by 
\be
 \zeta_{0\alpha}=\frac{\psi_{0\alpha}}{E}.  \label{rel_zeta-psi}
\ee
In fact, we can easily show that the right-hand-side of this equation 
transforms as $\zeta_{0\alpha}$ in Eq.(\ref{dltQ1}) 
under the $Q_1$-SUSY. 

From Eq.(\ref{rel_zeta-psi}), we can see that 
\be
 \psi_{0\alpha}=\zeta_{0\alpha}+\cO(\kp^2). 
\ee
This means that  
\be
 \vph_0=\rho_0+\cO(\kp^2),  
\ee
as mentioned before. 

Using $E$ defined by Eq.(\ref{def_E}), an invariant action for 
a matter field~$\phi$ can be written in the form of 
\be
 S_{\rm matter}=\int\dr^3x{\rm d}^2\tht_2 \:  
 E \cF(\phi,\cD_{2\alpha}\phi,\cD_m\phi,\cdots), 
\ee
where $\cF$ is a 3D Lorentz invariant function of $\phi$ and 
its covariant derivatives. 

From Eqs.(\ref{def_psi}) and (\ref{dlt_psi}), we can read off 
the $Q_1$-transformation law of $\vph_0$ as 
\be
 \dlt_{\xi_1}^{Q_1}\vph_0=-2\kp^{-1}\xi_1\tht_2+\kp\xi_1D_2\Xi. 
 \label{dlt_vph}
\ee
If we define a complex scalar superfield  
\be 
 \phi_0 \equiv \kp\vph_0+i\kp^2\Xi,  \label{def_phi0}
\ee
Eqs.(\ref{dlt_Xi}) and (\ref{dlt_vph}) are collectively written 
as\footnote{ 
By using $\phi_0$, we can construct a linear realization 
for partial SUSY breaking: 3D $\cN=2\to\cN=1$~\cite{ivanov2}. 
}
\be
 \dlt_{\xi_1}^{Q_1}\phi_0 = -2\xi_1\tht_2-i\xi_1D_2\phi_0. 
 \label{dltQ1phi}
\ee

The $P_2$-transformation law of $\phi_0$ can be defined as 
\be
 \dlt_a^{P_2}\phi_0=a. \label{dltZphi}
\ee
Eqs.(\ref{dltQ1phi}) and (\ref{dltZphi}) form 
the SUSY algebra~(\ref{SUSYalg1}).

\section{Mode-expansion approach} \label{MEA}
Now we will discuss the mode-expansion approach to obtain the effective 
action on the BPS wall. 
Here we will consider the 4D $\cN=1$ generalized Wess-Zumino model 
as a bulk theory. 
The action is 
\be
 S=\int\dr^4x{\rm d}^2\tht{\rm d}^2\btht \: K(\Phi,\bar{\Phi}) 
 +\int\dr^4x{\rm d}^2\tht \: W(\Phi)+\int\dr^4x{\rm d}^2\btht 
 \: \bar{W}(\bar{\Phi}), 
 \label{WZaction}
\ee
where $\Phi^i$ ($\bar{\Phi}^{\bar{i}}$) are (anti-) chiral superfields 
and 
\be
 \Phi^i(y,\tht) = A^i(y)+\sqrt{2}\tht\Psi^i(y)+\tht^2F^i(y). 
 \;\;\; (y^\mu \equiv x^\mu+i\tht\sgm^\mu\bar{\tht})
\ee

We assume that this theory has a BPS domain wall~$A^i=\Acl^i(x_2)$. 
The classical solution~$\Acl^i(x_2)$ satisfies the following BPS equation. 
\be
 \der_2 A^i = e^{i\dlt}K^{i\bar{j}}\bar{W}_{\bar{j}}, \label{BPSeq}
\ee
where $\der_2\equiv \der/\der x_2$, and $K^{i\bar{j}}$ is an inverse matrix 
of the K\"{a}hler metric 
$K_{i\bar{j}}\equiv \der^2 K/\der\Phi^i\der\bar{\Phi}^{\bar{j}}$. 
Lower indices denote derivatives in terms of corresponding superfields. 
$\dlt$ is a phase determined by $\Acl^i$, 
\be
 \dlt \equiv \arg\left(\int_\Gm{\rm d}W\right), 
\ee
where $\Gm$ is an orbit for $\Acl^i(x_2)$ on the target space of the scalar field. 

Now we decompose the Grassmannian coordinates~$\tht$ and $\btht$ as 
\be
 \tht^\alpha = \frac{e^{i\dlt/2}}{\sqrt{2}}(\tht_1^\alpha+i\tht_2^\alpha), \;\;\; 
 \btht^{\dot{\alpha}}=\frac{e^{-i\dlt/2}}{\sqrt{2}}(\tht_1^\alpha-i\tht_2^\alpha), 
\ee
where $\tht_i^\alpha$ ($i=1,2$) are 3D Majorana spinors, 
and corresponding to this, we decompose the supercharges as 
\be
 Q_\alpha=\frac{e^{-i\dlt/2}}{\sqrt{2}}(Q_{1\alpha}-iQ_{2\alpha}), \;\;\;
 \bar{Q}_{\dot{\alpha}}=-(Q_\alpha)^*=-\frac{e^{i\dlt/2}}{\sqrt{2}}
 (Q_{1\alpha}+iQ_{2\alpha}).  \label{Q-decomp}
\ee
Then it follows that 
\be
 \tht Q+\btht\bar{Q}=\tht_1Q_1+\tht_2Q_2. 
\ee
In this case, $Q_1$ and $Q_2$ correspond to the broken and 
the unbroken supercharges, respectively. 
Under the decomposition~(\ref{Q-decomp}), the 4D $\cN=1$ SUSY algebra 
is rewritten as Eq.(\ref{SUSYalg1}). 

After performing the integration in terms of $\tht_1$, 
the original action~(\ref{WZaction}) becomes~\cite{sakamura1} 
\be
 S=\int\dr^3x{\rm d}^2\tht_2{\rm d}x_2 
 \brc{K_{i\bar{j}}D_2^\alpha\vph^i D_{2\alpha}\bar{\vph}^{\bar{j}}
 -2iK_i\der_2\vph^i+4\Im\brkt{e^{-i\dlt}W(\vph)}}. 
 \label{S_for_vph}
\ee
Here 
\be
 \vph^i(x^m,x_2,\tht_2)\equiv e^{ix^mP_m+ix_2P_2+\tht_2Q_2}\times A^i(0), 
 \label{def_vph}
\ee
where the definition of the action of the generators 
on the fields~$\times$ is given in Appendix~\ref{act_on_fld}, 
and its relation to the 4D chiral superfield~$\Phi^i$ is 
\be
 \Phi^i(x^m,x_2+\tht_1\tht_2,\tht_1,\tht_2)
 =e^{-i\tht_1D_2+i\tht_1^2\der_2}\vph^i(x^m,x_2,\tht_2). 
\ee

Using $\vph^i$, we can rewrite the superfield equations of motion 
\be
 -\frac{1}{4}D^2K_i+W_i=0, 
\ee
as  
\be
 -\frac{i}{2}\brc{K_{i\bar{j}}D_2^2\bar{\vph}^{\bar{j}}
 +K_{i\bar{j}\bar{k}}D_2^\alpha \bar{\vph}^{\bar{j}}
 D_{2\alpha}\bar{\vph}^{\bar{k}}}
 -K_{i\bar{j}}\der_2\vph^{\bar{j}}+e^{-i\dlt}W_i=0. 
 \label{vphEOM}
\ee

Then, the equations of motion for the fluctuation field~$\tl{\vph}^i$ 
around the background~$\vph_{\rm cl}^i=\Acl^i(x_2)$ can be obtained 
by substituting $\vph^i=\vph_{\rm cl}^i+\tl{\vph}^i$ into Eq.(\ref{vphEOM}). 
Using the BPS equation~(\ref{BPSeq}), it can be written as 
\be
 -\frac{1}{2}K_{i\bar{j}}(\Acl)D_2^2\bar{\tl{\vph}}^{\bar{j}}
 +i\brc{\cD_y\bar{\tl{\vph}}_i-e^{-i\dlt}\cD_iW_j(\Acl)\tl{\vph}^j}+\cdots=0, 
 \label{linEOM}
\ee
where the ellipsis denotes higher terms for $\tl{\vph}$, and 
\be
 \bar{\tl{\vph}}_i\equiv K_{i\bar{j}}(\Acl)\bar{\tl{\vph}}^{\bar{j}}, \;\;\;  
 \cD_y\bar{\tl{\vph}}_i\equiv \der_2\bar{\tl{\vph}}_i
 -\Gm_{ij}^k(\Acl)\der_2\Acl^j\bar{\tl{\vph}}_k, \;\;\;
 \cD_iW_j\equiv W_{ij}-\Gm_{ij}^kW_k. 
\ee 
Here $\Gm_{ij}^k\equiv K^{k\bar{l}}K_{ij\bar{l}}$ is 
a connection on the K\"{a}hler manifold. 

From Eq.(\ref{linEOM}), we can find the mode equation, 
\be
 i\brc{\cD_y\bar{u}_{(n)i}-e^{-i\dlt}\cD_iW_j(\Acl)u_{(n)}^j}=m_{(n)}\bar{u}_{(n)i}, 
\ee
where $\bar{u}_{(n)i}\equiv K_{i\bar{j}}(\Acl)\bar{u}_{(n)}^{\bar{j}}$. 

Using the mode function~$u_{(n)}^i(x_2)$, we can expand the 4D field~$\vph^i$ as 
\be
 \vph^i(x^m,x_2,\tht_2)=\Acl^i(x_2)+\frac{1}{\sqrt{2}}\sum_{n=0}^\infty u_{(n)}^i(x_2)
 \vph_{(n)}(x^m,\tht_2).  \label{mode_ex_vph}
\ee
Here we have chosen the normalization of $u_{(n)}^i(x_2)$ as\footnote{
We have assumed eigenvalues~$m_{(n)}$ to be real. 
See Appendix~C in Ref.\cite{sakamura1}. 
}
\be
 \int\dr x_2 \: \Re\brc{\bar{u}_{(n)i}(x_2)u_{(m)}^i(x_2)}=\dlt_{nm}. 
\ee

By substituting Eq.(\ref{mode_ex_vph}) into Eq.(\ref{S_for_vph}) 
and performing the $x_2$-integration, 
the original 4D action~(\ref{WZaction}) can be rewritten 
in terms of infinite 3D superfields as follows. 
\be
 S=\int\dr^3x{\rm d}^2\tht_2 \: 
 \sbk{\sum_{n=0}^\infty\brc{\frac{1}{2}(D_2\vph_{(n)})^2+m_{(n)}\vph_{(n)}^2}
 +(\mbox{interaction terms})}.  \label{naive_S3}
\ee
Finally, by integrating out the heavy modes, 
we can derive the low-energy effective action on the domain wall.

\section{Modified mode-expansion} \label{mdfd_MEA}
\subsection{Nambu-Goldstone mode}
In the expression~(\ref{naive_S3}), the $Q_2$-SUSY is manifest 
since it is written in terms of superfields for the $Q_2$-SUSY. 
However, the nonlinear $Q_1$-SUSY is not unclear. 
In fact, the $Q_1$-transformation of each mode~$\vph_{(n)}$ defined by 
the mode-expansion~(\ref{mode_ex_vph}) does not close on $\vph_{(n)}$. 

Considering Eq.(\ref{mode_ex_vph}) and the $Q_1$-transformation 
of the 4D field~$\vph^i$, which is derived in Appendix~\ref{trf_vph},  
we can extract the $Q_1$-transformation law of $\vph_{(n)}$ as follows. 
\be
 \dlt_{\xi_1}^{Q_1}\vph_{(n)}=-2\kp^{-1}\xi_1\tht_2\dlt_{n,0}
 -2\xi_1\tht_2\sum_m U_{nm}\vph_{(m)}+\sum_mV_{nm}\xi_1D_2\vph_{(m)}, 
 \label{trfQ1vph}
\ee
where 
\bea
 \kp^{-2} \defa \int\dr x_2 \:\brc{K_{i\bar{j}}(\Acl)
 \der_2\Acl^i\der_2\bar{A}_{\rm cl}^{\bar{j}}
 +K^{i\bar{j}}(\Acl)W_i(\Acl)\bar{W}_{\bar{j}}(\bar{A}_{\rm cl})} \nonumber\\
 \eql 2\int\dr x_2 \: K_{i\bar{j}}(\Acl)
 \der_2\Acl^i\der_2\bar{A}_{\rm cl}^{\bar{j}}
\eea
is the tension of the domain wall, and 
\bea
 U_{nm} \defa \int\dr x_2 \: \Re\brc{\bar{u}_{(n)i}\der_2u_{(m)}^i}, 
 \nonumber\\
 V_{nm} \defa \int\dr x_2 \: \Im\brc{\bar{u}_{(n)i}u_{(m)}^i}. 
\eea
Here we have used the fact that 
\be
 \frac{1}{\sqrt{2}}u_{(0)}^i(x_2)=\kp\der_2\Acl^i(x_2). 
\ee

We can derive the $P_2$-transformation law of $\vph_{(n)}$ in a similar way.  
\be
 \dlt_a^{P_2}\vph_{(n)}=a\brkt{\kp^{-1}\dlt_{n,0}+\sum_mU_{nm}\vph_{(m)}}. 
 \label{trfZvph}
\ee

As mentioned above, either transformation~(\ref{trfQ1vph}) 
or (\ref{trfZvph}) does not close on each $\vph_{(n)}$. 

Next, we will modify the mode-expansion~(\ref{mode_ex_vph}) 
so that the broken symmetries~$Q_1$ and $P_2$ are nonlinearly realized 
on each mode. 
Consider the following mode-expansion. 
\be
 \vph^i(x^m,x_2,\tht_2)=\Acl^i(x_2+\phi_0)
 +\frac{1}{\sqrt{2}}\sum_{n\neq 0}u_{(n)}^i(x_2+\phi_0)\vph_{(n)}(x^m,\tht_2), 
 \label{mdfd_mode_ex1}
\ee
where $\phi_0$ is a {\it complex} scalar function of the NG mode~$\vph_0$. 
From this mode-expansion and Eqs.(\ref{Q1vph}) and (\ref{P2vph}) 
in Appendix~\ref{trf_vph}, 
the transformation laws of $\phi_0$ are read off as 
\bea
 \dlt_{\xi_1}^{Q_1}\phi_0 \eql -2\xi_1\tht_2-i\xi_1D_2\phi_0, \nonumber\\
 \dlt_a^{P_2}\phi_0 \eql a. 
\eea
These coincide with Eqs.(\ref{dltQ1phi}) and (\ref{dltZphi}). 
Therefore, $\phi_0$ defined by the modified 
mode-expansion~(\ref{mdfd_mode_ex1}) can be identified with $\phi_0$ 
in Eq.(\ref{def_phi0}) in the previous section. 

Indeed, by substituting Eq.(\ref{mdfd_mode_ex1}) into Eq.(\ref{S_for_vph}), 
dropping the massive modes~$\vph_{(n)}$ ($n\neq 0$),  
and carrying out the $x_2$-integration, 
we will reproduce the supersymmetric Nambu-Goto action~Eq.(\ref{Sng}).

\subsection{Matter action}
Although the NG mode~$\vph_0$ is identified with that of 
the nonlinear realization in the mode-expansion~(\ref{mdfd_mode_ex1}), 
the $Q_1$-transformation law of the other modes~$\vph_{(n)}$ ($n\neq 0$) 
does not close on themselves. 
Then, in this subsection, we will further modify the mode-expansion 
so that $\vph_{(n)}$ ($n\neq 0$) transforms in the standard nonlinear 
transformation under the $Q_1$-SUSY. 

Before we proceed, let us comment on the validity of such modification. 
From the 3D viewpoint, the modification of mode-expansion 
from Eq.(\ref{mode_ex_vph}) to Eq.(\ref{mdfd_mode_ex1}) 
corresponds to the redefinition of the superfield~$\vph_{(n)}$. 
Note that such field redefinition involves space-time derivatives. 
In fact, although the original theory~(\ref{WZaction}) contains 
no derivative couplings, the resulting effective theory has 
derivative couplings. 
(See Eq.(\ref{Sng}) or (\ref{Nambu-Goto}).) 
Hence, this field redefinition induces a new cut-off scale into 
the theory, which is $f\equiv \kp^{-2/3}$. 

Naively thinking, the mass of the first excited mode is thought to be $\cO(f)$, 
and thus all modes except the NG mode~$\vph_0$ in Eq.(\ref{mdfd_mode_ex1}) 
should be integrated out. 
However, there can be exist modes lighter than the cut-off scale~$f$ 
in some models. 
For example, the theory with 
\bea
 K(\Phi,\bar{\Phi},X,\bar{X}) \eql \Phi\bar{\Phi}+X\bar{X}, \nonumber\\
 W(\Phi,X) \eql \Lmd^2\Phi-\frac{g}{3}\Phi^3-h\Phi X^2, \;\;\;
 (\Lmd,g,h>0)
\eea
has the following BPS domain wall. 
\bea
 \Acl^\Phi(x_2) \eql \frac{\Lmd}{\sqrt{g}}\tanh(\sqrt{g}\Lmd x_2), \nonumber\\
 \Acl^X(x_2) \eql 0. 
\eea
In this case, the dynamical scale of the domain wall $f$ is 
\be
 f=\brkt{\frac{8}{3\sqrt{g}}}^{1/3}\Lmd, 
\ee
and a ``matter field'' $X$ contains modes with the mass 
eigenvalues~\cite{maru,hisano}
\be
 m_{(n)}=\sqrt{n\brkt{\frac{4h}{g}-n}}\sqrt{g}\Lmd. \;\;\;\;\;
 \brkt{n=0,1,2,\cdots < \frac{2h}{g}}
\ee
This means that in the case of $g\ll h<1$, which corresponds 
to the fat brane~\cite{kaplan}, 
there are many light modes   
which satisfy the condition~$m_{(n)}\ll f$. 
In the brane-world model-building, such light modes, especially the massless 
modes besides the NG mode, play important roles. 
Therefore, the further modification of the mode-expansion 
involving the ``matter'' modes~$\vph_{(n)}$ ($n\neq 0$) is a useful work. 

The modified mode-expansion is 
\be
 \vph^i=\Acl^i(x_2+\phi_0)+\frac{1}{\sqrt{2}}
 \sum_{n\neq 0}u_{(n)}^i(x_2+\phi_0)\tl{\vph}_{(n)}(x^m,\tht_2), 
 \label{mdfd_mode_ex2}
\ee
with 
\be
 \tl{\vph}_{(n)}\equiv \vph_{(n)}-i\kp\zeta_0D_2\vph_{(n)}
 -\kp^2\zeta_0^\alpha D_{2\alpha}\zeta_0^\beta D_{2\beta}\vph_{(n)}
 +\frac{\kp^2}{4}\zeta_0^2D_2^2\vph_{(n)}
 +\cO(\kp^3).  \label{def_tlvph}
\ee
Here $\zeta_0$ is defined by Eq.(\ref{rel_zeta-psi}). 
In this case, each mode transforms under the broken symmetries as follows. 
\bea
 \dlt_a^{P_2}\vph_{(n)} \eql 0, \nonumber\\
 \dlt_{\xi_1}^{Q_1}\vph_{(n)} \eql 
 -i\kp\xi_1\gm_{(3)}^m\zeta_0\der_m\vph_{(n)}+\cO(\kp^2). 
 \label{ap_SNL}
\eea
These coincide with Eqs.(\ref{dltZ}) and (\ref{dltQ1}) up to $\cO(\kp)$. 

In fact, by substituting Eq.(\ref{mdfd_mode_ex2}) into Eq.(\ref{S_for_vph}), 
we can obtain the following Lagrangian after somewhat tedious 
calculation\footnote{
For simplicity, we have assumed the minimal K\"{a}hler potential. 
}. 
\bea
 \cL^{(3)} \eql \int\dr^2\tht_2 \: 2\Xi +\int\dr^2\tht_2 \:
 E\left[\sum_{n\neq 0}\brc{\frac{1}{2}\brkt{\cD_2\vph_{(n)}}^2
 +m_{(n)}\vph_{(n)}^2}+V(\vph_{(n)}) \right. \nonumber\\
 && +\kp\sum_{n,m}g_{(nm)}\vph_{(m)}
 \brkt{\cD_2^2\zeta_0\cD_2\vph_{(n)}+2i\cD_2\gm_{(3)}^m\zeta_0
 \cD_m\vph_{(n)}}+\kp\cD_2^\alpha\zeta_{0\alpha}U(\vph_{(n)}) \nonumber\\
 && \left. +\kp^2\brkt{\frac{1}{2}\cD_2^\alpha\zeta_0^\beta\cD_{2\alpha}
 \zeta_{0\beta}+\cD_2^\alpha\zeta_0^\beta\cD_{2\beta}\zeta_{0\alpha}}
 \brkt{\cD_2\vph_{(n)}}^2+\kp^2\cD_2^\alpha\zeta_0^\beta\cD_{2\beta}
 \zeta_{0\alpha}\tl{V}(\vph_{(n)})\right] \nonumber\\
 && +\cO(\kp^3), 
 \label{3D_L}
\eea
where $E$ is defined by Eq.(\ref{def_E}), and 
\bea
 g_{(nm)} \defa \int\dr x_2 \: \Im\brkt{\bar{u}_{(n)i}u_{(m)}^i}, \nonumber\\
 V \defa \sum_{N=3}^\infty \sum_{n_1,\cdots,n_N} 
 \brc{4\int\dr x_2 \: \Im\brkt{\frac{e^{-i\dlt}}{N!}u_{(n_1)}^{i_1}\cdots 
 u_{(n_N)}^{i_N}W_{i_1\cdots i_N}}}\vph_{(n_1)}\cdots\vph_{(n_N)}, 
 \nonumber\\
 \tl{V} \defa \sum_{N=1}^\infty \sum_{n_1,\cdots,n_N} 
 \brc{4\int\dr x_2 \: \Im\brkt{\frac{e^{-i\dlt}}{N!}u_{(n_1)}^{i_1}\cdots 
 u_{(n_N)}^{i_N}W_{i_1\cdots i_N}}}\vph_{(n_1)}\cdots\vph_{(n_N)}, 
 \nonumber\\
 U \defa \sum_{N=2}^\infty \sum_{n_1,\cdots,n_N} 
 \brc{4\int\dr x_2 \: \Re\brkt{\frac{e^{-i\dlt}}{N!}u_{(n_1)}^{i_1}\cdots 
 u_{(n_N)}^{i_N}W_{i_1\cdots i_N}}}\vph_{(n_1)}\cdots\vph_{(n_N)}.  
\eea

The above Lagrangian certainly has an invariance under 
the $Q_1$-SUSY, which is nonlinearly realized. 
The first term is the supersymmetric Nambu-Goto Lagrangian, 
and the remaining part corresponds to the matter Lagrangian.

\section{Broken Lorentz and $U(1)_R$ symmetries} \label{bk_Lorentz}
In the matter Lagrangian in Eq.(\ref{3D_L}), note that 
the NG mode~$\vph_0$ appears not only through $E$ and $\cD_2$, 
but also in the form such as $\cD_2^\alpha\zeta_{0\beta}$. 
This means that even if we know the Lagrangian in the limit of $\kp\to 0$ 
in some way, we cannot reproduce the full Lagrangian~Eq.(\ref{3D_L}) 
by using the method of the nonlinear realization. 
This stems from the fact that we did not respect the broken Lorentz 
symmetry and the $U(1)_R$ symmetry in the modification of 
the mode-expansion~(\ref{mdfd_mode_ex2}) and (\ref{def_tlvph}). 

In the nonlinear realization for space-time symmetries, 
if we take into account only the (super-)translational generators 
as generators of the whole bulk symmetry~$G$,  
there will be an ambiguity of inserting a dimensionless tensor, 
such as $\kp\cD_2^\alpha\zeta_{0\beta}$, into the $G$-invariant effective 
action \cite{bagger1}. 
For example, let us assume that the Lagrangian at $\kp\to 0$ is\footnote{
In fact, such Lagrangian can be calculated much easier than Eq.(\ref{3D_L}) 
since Eqs.(\ref{mdfd_mode_ex2}) and (\ref{def_tlvph}) 
become very simple forms 
by dropping $\vph_0$ (and thus $\zeta_0$) from Eq.(\ref{mdfd_mode_ex2}). 
} 
\be
 \cL^{(3)}_{\kp\to 0} = \int\dr^2\tht_2 \: 
 \sum_{n\neq 0}\brc{\frac{1}{2}\brkt{D_2\vph_{(n)}}^2+m_{(n)}\vph_{(n)}^2}
 +V(\vph_{(n)}). 
\ee
Then, the most general Lagrangian that is $P_2$- and $Q_1$-invariant is 
written as 
\bea
 \cL^{(3)} \eql \int\dr^2\tht_2 \: E \left[\sum_{n\neq 0}
 \brc{\frac{1}{2}\brkt{\cD_2\vph_{(n)}}^2+m_{(n)}\vph_{(n)}^2}+V(\vph_{(n)})
 \right. \nonumber\\
 &&\left.+\cF(\cD_2^\alpha\zeta_{0\beta},\cD_m\zeta_{0\alpha},
 \cD_2^\alpha\cD_2^\beta\zeta_{0\gm},\cdots)\right], 
 \label{gen_L}
\eea
where $\cF$ denotes 3D Lorentz-invariant terms that include at least one 
covariant derivative of $\zeta_0$. 
Note that the above Lagrangian has an ambiguity of adding terms 
including arbitrary numbers of $\kp\cD_2^\alpha\zeta_{0\beta}$ 
since such tensor is dimensionless. 
In fact, the second and the third lines in Eq.(\ref{3D_L}) correspond to 
$\cF$ in Eq.(\ref{gen_L}), and cannot be determined by 
the nonlinear realization. 

The above ambiguity already exists in the definition of 
$\tl{\vph}_{(n)}$ in Eq.(\ref{def_tlvph}). 
We may add terms, such as $\kp\cD_2^\alpha\zeta_{0\alpha}\vph_{(n)}$, 
to the definition of $\tl{\vph}_{(n)}$ since the resulting transformation 
laws of $\vph_{(n)}$ do not change. 
Such terms contribute to $\cF$ in Eq.(\ref{gen_L}). 

In order to remove such ambiguity, 
we should take the full symmetry group~$G$ as  
\be
 G_{\rm max}=\brc{P_m,P_2,Q_{1\alpha},Q_{2\alpha},M_{mn},K_m,R}, 
\ee
which contains the maximal automorphism group of the 4D $\cN=1$ 
SUSY algebra. 
Here $M_{mn}$ and $K_m\equiv M_{m2}$ denote the 4D Lorentz generators, 
and $R$ is a generator of $U(1)_R$.  
In this case, the vacuum stability subgroup is 
\be
 H_{\rm max}=\brc{P_m,Q_{2\alpha},M_{mn}}. 
\ee
The algebra among these generators is listed in Eq.(\ref{SP_alg}) 
in Appendix~\ref{superPalg}. 

The transformation laws of $\phi_0$ in the nonlinear realization 
under the $K$- and $R$-transformations can be defined by  
\bea
 \dlt_v^K\phi_0 \eql v^m\brkt{-x_m-\phi_0\der_m\phi_0
 +\frac{1}{2}\tht_2\gm_{(3)m}D_2\phi_0}, \label{dltKphi}\\
 \dlt_r^R\phi_0 \eql r\brkt{\tht_2^2+i\tht_2D_2\phi_0}, 
 \label{dltRphi}
\eea
where $v^m$ and $r$ are transformation parameters. 
These form the SUSY algebra~(\ref{SP_alg}) 
together with Eqs.(\ref{dltQ1phi}) and (\ref{dltZphi}). 

From Eq.(\ref{dltKphi}), 
\be
 \dlt_v^K \Xi=v^m\brc{-\kp\der_m\brkt{\vph_0\Xi}
 +\frac{i}{2}\kp^{-1}D_2^\alpha\brkt{(\gm_{(3)m}\tht_2)_\alpha\vph_0}}. 
\ee
This is a total derivative and thus 
the NG action~$S_{\rm NG}$ in Eq.(\ref{Sng}) is invariant 
under the $K$-transformation. 

For the $U(1)_R$ symmetry, on the other hand, 
we can see from Eq.(\ref{dltRphi}), 
\be
 \dlt_r^R\Xi=\kp^{-1}r\tht_2D_2\vph_0. 
\ee
This means 
\be
 \dlt_r^R S_{\rm NG}=\kp^{-1}r\int\dr^3x \: 2f_0, 
\ee
where $f_0$ is the auxiliary field of $\vph_0$. 
So $S_{\rm NG}$ does not have the off-shell $U(1)_R$ symmetry. 
However, considering the fact that $f_0=0$ on shell 
as mentioned in Appendix~\ref{SUSY_NG}, 
we can see that $S_{\rm NG}$ is $R$-invariant {\it on shell}. 

As a result, $S_{\rm NG}$ obtained in the nonlinear realization 
is invariant under the full $G_{\rm max}$ symmetry. 

On the other hand, the NG action obtained by 
the mode-expansion~(\ref{mdfd_mode_ex2}) does not have 
an invariance under the broken Lorentz symmetry. 
Although it has the same form as Eq.(\ref{Sng}), 
the $K$-transformation of $\phi_0$ does not coincide with Eq.(\ref{dltKphi}). 
Unlike the nonlinear realization, the transformation laws of each mode 
are determined from those of the bulk theory in the mode-expansion approach. 
The $K$-transformation law of $\vph^i$ is listed in Eq.(\ref{Kvph}) 
in Appendix~\ref{trf_vph}. 
Due to the explicit appearance of $x_2$ in Eq.(\ref{Kvph}),   
the $K$-transformation of $\phi_0$ does not close on $\phi_0$. 

Now we will try to modify 
the mode-expansion~(\ref{mdfd_mode_ex2}) further, 
so that $\phi_0$ transforms properly under all broken 
symmetries~$G_{\rm max}/H_{\rm max}$. 

We propose the following mode-expansion. 
\be
 \vph^i(x^m,x_2,\tht_2) = \Acl^i(\hat{x}_2)
 +\frac{1}{\sqrt{2}}\sum_{n\neq 0}u_{(n)}^i(\hat{x}_2)
 \tl{\vph}_{(n)}(\hat{x}^m,\tht_2),  \label{mdfd_mode_ex3}
\ee
where
\bea
 \hat{x}^m \defa x^m-x_2\der^m\phi_0-\frac{\kp^2}{2}\der^m(\vph_0^2)
 +\frac{\kp^2}{2}x_2\der^m(\der^n\vph_0\der_n\vph_0)+\cO(\kp^3), 
 \nonumber\\
 \hat{x}_2 \defa x_2+\phi_0-\frac{\kp^2}{2}x_2 \der^n\vph_0\der_n\vph_0
 +\cO(\kp^3),  \label{def_hat_cdnt}
\eea
and $\tl{\vph}_{(n)}$ is the one defined in Eq.(\ref{def_tlvph}). 

With this mode-expansion and Eq.(\ref{Kvph}), we can derive 
the $K$-transformation of each mode. 
\bea
 \dlt_v^K\phi_0 \eql v^m\brkt{-x_m+\frac{1}{2}\tht_2\gm_{(3)m}D_2\phi_0}
 +\cO(\kp^2),  \label{dltK_phi0}\\
 \dlt_v^K\tl{\vph}_{(n)} \eql v^m\brkt{-\phi_0\der_m\tl{\vph}_{(n)}
 +\frac{1}{2}\tht_2\gm_{(3)m}D_2\tl{\vph}_{(n)}}+\cO(\kp^2). 
 \label{dltK_tlvph}
\eea
Eq.(\ref{dltK_phi0}) coincides with Eq.(\ref{dltKphi}) up to $\cO(\kp)$. 
Considering the definition of $\tl{\vph}_{(n)}$ in Eq.(\ref{def_tlvph}), 
Eq.(\ref{dltK_tlvph}) is translated into  
\be
 \dlt_v^K\vph_{(n)}=v^m\brc{-\kp\vph_0\der_m\vph_{(n)}
 +\frac{\kp}{2}\zeta_0\gm_{(3)m}\tht_2\der_m\vph_{(n)}
 +\frac{i}{2}\kp\zeta_0\gm_{(3)m}D_2\vph_{(n)}}+\cO(\kp^2). 
 \label{dltK_vph}
\ee
Noticing that $\vph_0=\rho_0+\cO(\kp^2)$, this coincides 
with the standard nonlinear transformation~Eq.(\ref{dltK_st}) 
in Appendix~\ref{SNT}. 

In a similar way, we can derive the transformation laws 
for other symmetries. 

For the broken SUSY, we obtain 
\bea
 \dlt_{\xi_1}^{Q_1}\phi_0 \eql -2\xi_1\tht_2-i\xi_1D_2\phi_0+\cO(\kp^2), 
 \\
 \dlt_{\xi_1}^{Q_1}\vph_{(n)} \eql -i\kp\xi_1\gm_{(3)}^m\zeta_0
 \der_m\vph_{(n)}+\cO(\kp^2).  \label{dltQ1_vph}
\eea

For the $U(1)_R$ symmetry, by using Eq.(\ref{Rvph}) in Appendix~\ref{trf_vph}, 
we obtain 
\bea
 \dlt_r^R\phi_0 \eql r(\tht_2^2+i\tht_2D_2\phi_0)+\cO(\kp^2), \\
 \dlt_r^R\vph_{(n)} \eql -i\kp r\zeta_0\gm_{(3)}^m\tht_2\der_m\vph_{(n)}
 -\kp r\zeta_0D_2\vph_{(n)}+\cO(\kp^2). \label{dltR_vph}
\eea

Thus, $\phi_0$ transforms in the desired way in the case of our modified 
mode-expansion~(\ref{mdfd_mode_ex3}) and (\ref{def_hat_cdnt}). 
Note that Eqs.(\ref{dltQ1_vph}) and (\ref{dltR_vph}) are 
the standard nonlinear transformations, which are listed in 
Eq.(\ref{dltQ1_st}) and (\ref{dltR_st}) in Appendix~\ref{SNT}. 
Therefore, the modified mode-expansion~(\ref{mdfd_mode_ex3}) and (\ref{def_hat_cdnt}) 
defines modes on which all the broken symmetries are nonlinearly realized 
at least up to $\cO(\kp)$.

\section{Summary and discussion} \label{summary}
There are mainly two different approaches to derive the low-energy 
effective theory in the background of a BPS domain wall. 
Each approach has its own advantages and disadvantages. 

The first one is the nonlinear realization approach. 
We can construct an effective action on the wall 
that is invariant under both the broken and the unbroken SUSYs 
by using the nonlinear realization technique. 
This approach is useful when we discuss the general properties 
of BPS brane-like objects, such as BPS walls or supermembranes. 
However, in this approach, we neglect the wall width and 
cannot discuss the specific wall profile. 
Namely, we cannot determine parameters 
of the effective theory by this approach. 

The second one is the mode-expansion approach. 
In this approach, we start directly from the bulk theory. 
So we need to specify the bulk theory and thus 
the results are model-dependent. 
Since we explicitly derive the effective theory from the bulk theory, 
the relation between the bulk 4D superfields and the 3D superfields 
in the effective theory is clear in this approach. 
On the other hand, SUSY broken by the wall is not respected  
since its transformation of each mode does not close on itself. 

Therefore, it is very useful and instructive to clarify the relation 
between the above two approaches. 

In this paper, we proposed a modified mode-expansion so that the broken 
SUSY is nonlinearly realized on each mode. 
Indeed, our mode-expansion leads to 
a supersymmetric Nambu-Goto action for the NG mode, 
and to a matter action with a form expected from the nonlinear realization 
for the other modes. 
In particular, the NG mode~$\vph_0$ defined by our mode-expansion 
can be identified with that of the nonlinear realization 
{\it for all orders} in $\kp$. 
For the other modes, their transformation law under the broken SUSY 
coincides with the standard nonlinear transformation up to $\cO(\kp)$. 

Note that our modification of the mode-expansion corresponds to 
the redefinition of the 3D superfields. 
This field redefinition involves space-time derivatives and 
the scale~$f\equiv\kp^{-2/3}$. 
So the cut-off scale of the effective theory becomes $f$ after 
the field redefinition. 

We also showed that it is possible to modify the mode-expansion 
so that each mode transforms in the standard nonlinear transformation 
under not only the broken SUSY but also the broken Lorentz 
and $U(1)_R$ symmetries at least up to $\cO(\kp)$. 
However, whether the extension to higher orders in $\kp$ is possible 
is not clear to us. 

In this paper, we have discussed the BPS domain wall. 
When we construct a realistic brane-world model in a SUSY theory, 
we must consider the SUSY breaking mechanism. 
One of the simplest mechanism of SUSY breaking is the coexistence of 
the BPS and anti-BPS walls \cite{maru}. 
In such a case, each domain wall preserves an opposite half 
of the bulk SUSY, and all of the supersymmetries are broken 
in the whole system. 
In the thin wall limit, this corresponds to the one called 
the {\it pseudo-supersymmetry} \cite{klein}.  
The author of Ref.\cite{klein} derives the effective theory 
of the brane-antibrane system by using the nonlinear realization 
technique. 
In this case, the SUSY breaking effects are induced at loop level 
because tree-level couplings between the branes are absent. 
For the wall-antiwall system with a finite wall-width, on the other hand, 
SUSY breaking appears at tree level although its effects are 
exponentially suppressed for the distance between the walls \cite{maru}. 
To discuss the phenomenological arguments, it is useful 
to describe the effective theory on the wall in terms of 
the superfields and the SUSY breaking terms. 
Combining the method in Ref.\cite{klein} and the result of this paper, 
we can derive the effective theory in the wall-antiwall system 
in terms of the 3D superfields and the SUSY breaking terms. 
This work is now in progress. 

%

\vspace{5mm}

\begin{center}
{\bf Acknowledgments}
\end{center}
The author thanks the Yukawa Institute for Theoretical Physics at 
Kyoto University, where this work was initiated during 
the YITP-W-99-99 on ``Extra dimensions and Braneworld''. 
The author also thanks Koji Hashimoto for useful discussion.

\appendix
\section{Notations}
Basically, we follow the notations of Ref.\cite{WessBagger} 
for the 4D bulk theory. 
The notations for the 3D theories are as follows. 

We take the space-time metric as 
\be
 \eta^{mn}=\diag(-1,+1,+1). 
\ee

The 3D $\gm$-matrices, $(\gm_{(3)}^m)_\alpha^{\;\beta}$, 
can be written by the Pauli matrices as 
\be
 \gm_{(3)}^0=\sgm^2, \;\;\;
 \gm_{(3)}^1=-i\sgm^3, \;\;\;
 \gm_{(3)}^3=i\sgm^1, 
\ee
and these satisfy the 3D Clifford algebra, 
\be
 \brc{\gm_{(3)}^m,\gm_{(3)}^n}=-2\eta^{mn}. 
\ee

The spinor indices are raised and lowered by multiplying 
$\sgm^2$ from the left. 
\be
 \psi_\alpha=(\sgm^2)_{\alpha\beta}\psi^\beta, \;\;\;
 \psi^\alpha=(\sgm^2)^{\alpha\beta}\psi_\beta. 
\ee
We take the following convention of the contraction of spinor 
indices. 
\be
 \psi_1\psi_2\equiv \psi_1^\alpha\psi_{2\alpha}
 =(\sgm^2)_{\alpha\beta}\psi_1^\alpha\psi_2^\beta=\psi_2\psi_1. 
\ee

\subsection{Covariant derivatives} \label{cov_ders}
The algebra of the 3D $\cN=1$ SUSY preserved by the wall is 
\be
 \brc{Q_{2\alpha},Q_{2\beta}}=2(\gm_{(3)}^m\sgm^2)_{\alpha\beta}P_m. 
\ee
The representation of the generators on the 3D $\cN=1$ superspace 
$(x^m,\tht_2)$ is 
\bea
 \hat{P}_m \eql -i\der_m, \nonumber\\
 \hat{Q}_{2\alpha} \eql \der_{2\alpha}+i(\gm_{(3)}^m\tht_2)_\alpha\der_m. 
\eea

The SUSY covariant derivative for $\tht_2$ is  
\be
 D_{2\alpha}\equiv \der_{2\alpha}-i(\gm_{(3)}^m\tht_2)_\alpha\der_m. 
\ee

We list the covariant derivatives for superspace coordinates 
in the presence of the NG superfields as follows. 
They can be obtained by calculating 
the Cartan one-form~$\Omega^{-1}{\rm d}\Omega$ where $\Omega$ 
is defined in Eq.(\ref{Omega}). 
\bea
 \cD_m \rho_0 \eql (\omega^{-1})_m^{\;n}\der_n\rho_0=\der_m\rho_0+\cO(\kp^2), 
 \nonumber\\
 \cD_{2\alpha}\rho_0 \eql D_{2\alpha}\rho_0+2\zeta_{0\alpha}
 -i\kp^2D_{2\alpha}\zeta_0\gm_{(3)}^m\zeta_0(\omega^{-1})_m^{\;n}\der_n\rho_0, 
 \nonumber\\
 \cD_m\zeta_{0\beta} \eql (\omega^{-1})_m^{\;n}\der_n\zeta_{0\beta}
 =\der_m\zeta_{0\beta}+\cO(\kp^2), \nonumber\\
 \cD_{2\alpha}\zeta_{0\beta} \eql D_{2\alpha}\zeta_{0\beta}
 -i\kp^2D_{2\alpha}\zeta_0\gm_{(3)}^m\zeta_0(\omega^{-1})_m^{\;n}\der_n\zeta_{0\beta}, 
 \nonumber\\
 \cD_m\phi \eql (\omega^{-1})_m^{\;n}\der_n\phi=\der_m\phi+\cO(\kp^2), 
 \nonumber\\
 \cD_{2\alpha}\phi \eql D_{2\alpha}\phi
 -i\kp^2D_{2\alpha}\zeta_0\gm_{(3)}^m\zeta_0(\omega^{-1})_m^{\;n}\der_n\phi, 
 \label{cov_drvs1}
\eea
where $\phi$ denotes a matter field, and 
\be
 \omega_m^{\;n}\equiv \dlt_m^{\;n}+i\kp^2\der_m\zeta_0\gm_{(3)}^n\zeta_0. 
\ee

\subsection{Action of the generators on the fields} \label{act_on_fld}
The SUSY transformation~$\dlt^Q_\xi$ of a chiral 
supermultiplet~$(A,\Psi^\alpha,F)$ is defined by 
\bea
 \dlt^Q_\xi A \eql \sqrt{2}\xi\Psi, \nonumber\\
 \dlt^Q_\xi \Psi_\alpha \eql i\sqrt{2}(\sgm^\mu\bar{\xi})_\alpha
  \der_\mu A+\sqrt{2}\xi_\alpha F, \nonumber\\
 \dlt^Q_\xi F \eql i\sqrt{2}\bar{\xi}\bar{\sgm}^\mu\der_{\mu}\Psi. 
\eea

We define an action of the generators~$P_\mu$, $Q_\alpha$ and 
$\bar{Q}^{\dot{\alpha}}$ on the fields $\phi=A,\Psi^\alpha,F$ as 
\bea
 P_\mu \times \phi \defa -i\der_\mu\phi, \nonumber\\
 (\xi Q+\bar{\xi}\bar{Q})\times \phi \defa \dlt^Q_\xi \phi. 
\eea

Under the above definition, the chiral superfield~$\Phi$ can be 
written as 
\be
 \Phi(x,\tht,\btht)=e^{ix^\mu P_\mu+\tht Q+\btht\bar{Q}}\times A(0). 
\ee

There is a useful formula that converts covariant derivatives~$\cD_M$ 
into the corresponding generators~$\Gm_M$. 
\be
 \cD_M e^{iX^N \Gm_N} = e^{iX^N \Gm_N} \Gm_M, 
 \label{usf_fml}
\ee
where $X^M$ are coordinates for $\Gm_M$. 

For the SUSY generator~$Q_2$, for instance, 
\be
 \cD_{2\alpha} e^{ix^m P_m+\tht_2 Q_2} = e^{ix^m P_m+\tht_2 Q_2}
 Q_{2\alpha}. 
\ee

Using this formula, we can rewrite the chiral condition for $\Phi$  
as follows. 
\bea
 \bar{D}_{\dot{\alpha}}\Phi \eql 
 \bar{D}_{\dot{\alpha}}e^{\tht Q+\btht\bar{Q}}
 \times A(x) \nonumber\\
 \eql e^{\tht Q+\btht\bar{Q}}\bar{Q}_{\dot{\alpha}}\times A(x)=0. 
\eea
Then, 
\be
 \bar{Q}_{\dot{\alpha}}\times A(x)=
 -\frac{e^{i\dlt/2}}{\sqrt{2}}(Q_{1\alpha}+iQ_{2\alpha})\times A(x)
 =0. 
\ee
Namely, 
\be
 Q_{1\alpha}\times A = -iQ_{2\alpha}\times A.  \label{chiral_cond}
\ee

\section{Bosonic part of $S_{\rm NG}$} \label{SUSY_NG}
Here we will provide a brief derivation of the Nambu-Goto action 
from the effective action~$S_{\rm NG}$ defined in Eq.(\ref{Sng}). 
We denote the component fields of $\vph_0$ as 
\be
 \vph_0=a_0+\tht_2\eta_0+\frac{1}{2}\tht_2f_0. 
 \label{comp_vph}
\ee
In order to concentrate on the bosonic part of $S_{\rm NG}$, 
we will neglect the fermionic component~$\eta_0$ 
in the following. 
Then, 
\be
 \psi_0^2|_{\rm bosonic}=\frac{1}{4}\tht_2^2(f_0^2-\der^ma_0\der_ma_0), 
\ee
and 
\be
 D_2^2(\psi_0^2)|_{\rm bosonic}=-f_0^2+\der^ma_0\der_ma_0. 
\ee
Therefore, the auxiliary field~$f_0$ enters in $S_{\rm NG}$ 
only in a bilinear way. 
This means that an equation of motion for $f_0$ is $f_0$=0. 
Hence, after the elimination of $f_0$, the bosonic part of $S_{\rm NG}$ 
becomes  
\bea
 S_{\mbox{\scriptsize NG bosonic}}^{\mbox{\scriptsize on-shell}} \eql 
 \int\dr^3x 
 \frac{-\der^ma_0\der_ma_0}{1+\sqrt{1+\kp^2\der^ma_0\der_ma_0}} \nonumber\\
 \eql \int\dr^3x \: \kp^{-2}\brc{1-\sqrt{1+\kp^2\der^ma_0\der_ma_0}}. 
 \label{Nambu-Goto}
\eea
This is the Nambu-Goto action in the static gauge, as expected.

\section{Transformation of $\vph^i$ under the broken symmetries} 
\label{trf_vph}
The transformation laws of $\vph^i$ for the broken symmetries are obtained 
from the definition of $\vph^i$ in Eq.(\ref{def_vph}) 
and the SUSY algebra~(\ref{SUSYalg1}). 

For example, for the $Q_1$-transformation, 
\bea
 \dlt_{\xi_1}^{Q_1}\vph^i \eql \xi_1Q_1\times\vph^i
 =\xi_1Q_1e^{ix^mP_m+ix_2P_2+\tht_2Q_2}\times A^i(0) \nonumber\\
 \eql e^{ix^mP_m+ix_2P_2+\tht_2Q_2}(-2i\xi_1\tht_2P_2+\xi_1Q_1)\times A^i(0) 
 \nonumber\\
 \eql e^{ix^mP_m+ix_2P_2+\tht_2Q_2}(-2i\xi_1\tht_2P_2-i\xi_1Q_2)\times 
 A^i(0) \nonumber\\
 \eql (-2\xi_1\tht_2\der_2-i\xi_1D_2)\vph^i.  \label{Q1vph}
\eea
Here we have used the chiral condition~Eq.(\ref{chiral_cond})
and the formula~Eq.(\ref{usf_fml}). 

Similarly, for the $P_2$-transformation, 
\be
 \dlt_a^{P_2}\vph^i=iaP_2\times \vph^i=a\der_2\vph^i. \label{P2vph}
\ee

For the broken Lorentz transformation, 
\be
 \dlt_v^K\vph^i = iv^mK_m\times \vph^i
 =v^m\brc{x_2\der_m\vph^i-x_m\der_2\vph^i
 +\frac{1}{2}\tht_2\gm_{(3)m}D_2\vph^i}.  \label{Kvph}
\ee

For the $U(1)_R$ transformation, 
\be
 \dlt_r^R\vph^i = irR\times \vph^i 
 =r\brc{\tht_2^2\der_2\vph^i+i\tht_2D_2\vph^i}. \label{Rvph}
\ee

\section{In the case of $G=G_{\rm max}$}
Here we collect the main results of the nonlinear realization 
in the case that the full symmetry group $G$ is taken to be 
$G_{\rm max}=\brc{P_m,P_2,Q_{1\alpha},Q_{2\alpha},M_{mn},K_m,R}$. 
The vacuum stability subgroup is 
$H_{\rm max}=\brc{P_m,Q_{2\alpha},M_{mn}}$ in this case. 

\subsection{Super-Poincar\'{e} algebra} \label{superPalg}
The 3D $\cN=2$ super-Poincar\'{e} algebra with central extension is 
as follows. 

\bea
 \{Q_{1\alpha},Q_{1\beta}\} \eql \{Q_{2\alpha},Q_{2\beta}\}
 =2(\gm_{(3)}^m\sgm^2)_{\alpha\beta}P_m, \nonumber\\
 \{Q_{1\alpha},Q_{2\beta}\} \eql -\{Q_{2\alpha},Q_{1\beta}\}
 =2i(\sgm^2)_{\alpha\beta}P_2,  \nonumber\\
 \sbk{M_{mn},Q_{1\alpha}} \eql -i\brkt{\gm_{(3)mn}Q_1}_\alpha, \;\;\;
 \sbk{M_{mn},Q_{2\alpha}}=-i\brkt{\gm_{(3)mn}Q_2}_\alpha, \nonumber\\
 \sbk{K_m,Q_{1\alpha}} \eql -\frac{1}{2}\brkt{\gm_{(3)m}Q_2}_\alpha, \;\;\;
 \sbk{K_m,Q_{2\alpha}}=\frac{1}{2}\brkt{\gm_{(3)m}Q_1}_\alpha, \nonumber\\
 \sbk{R,Q_{1\alpha}} \eql -iQ_{2\alpha}, \;\;\;
 \sbk{R,Q_{2\alpha}}=iQ_{1\alpha}, \nonumber\\
 \sbk{M_{mn},P_l} \eql i(\eta_{lm}P_n-\eta_{ln}P_m), \;\;\;
 \sbk{M_{mn},P_2} = 0, \nonumber\\
 \sbk{K_m,P_n} \eql i\eta_{mn}P_2, \;\;\;
 \sbk{K_m,P_2} = -iP_m, \nonumber\\
 \sbk{M_{mn},M_{lp}} \eql 
 i(\eta_{ml}M_{np}-\eta_{nl}M_{mp}-\eta_{pn}M_{lm}+\eta_{pm}M_{ln}), 
 \nonumber\\
 \sbk{M_{mn},K_l} \eql i(\eta_{ml}K_n-\eta_{nl}K_m), \nonumber\\
 \sbk{K_m,K_n} \eql iM_{mn}.  \label{SP_alg}
\eea

\subsection{Standard nonlinear transformations} \label{SNT}
A coset element~$\hat{\Omega}$ can be parameterized as 
\be 
 \hat{\Omega}=e^{ix^mP_m+\tht_2Q_2}e^{i\rho_0P_2+\zeta_0Q_1}
 e^{i\lmd_0^mK_m+i\sgm_0 R},  \label{hatOmega}
\ee
where $\rho_0$, $\zeta_{0\alpha}$, $\lmd_0^m$ and $\sgm_0$ 
are the NG superfields for the corresponding generators. 

The transformation laws of each superfield are obtained 
by multiplying $\hat{\Omega}$ by corresponding group elements 
from the left. 

For $P_2$-transformation, 
\bea
 \dlt_a^{P_2}\rho_0 \eql \kp^{-1}a, \nonumber\\
 \dlt_a^{P_2}\zeta_{0\alpha} \eql \dlt_a^{P_2}\lmd_0^m 
 =\dlt_a^{P_2}\sgm_0 =\dlt_a^{P_2}\phi=0. 
 \label{dltP2_st}
\eea

For the broken SUSY, 
\bea
 \dlt_{\xi_1}^{Q_1}\rho_0 \eql -2\kp^{-1}\xi_1\tht_2
 -i\kp\xi_1\gm_{(3)}^m\zeta_0\der_m\rho_0, \nonumber\\
 \dlt_{\xi_1}^{Q_1}\zeta_{0\alpha} \eql \kp^{-1}\xi_{1\alpha} 
 -i\kp\xi_1\gm_{(3)}^m\zeta_0\der_m\zeta_{0\alpha}, \nonumber\\
 \dlt_{\xi_1}^{Q_1}\lmd_0^m \eql 
 =-i\kp\xi_1\gm_{(3)}^n\zeta_0\der_n\lmd_0^m, \nonumber\\
 \dlt_{\xi_1}^{Q_1}\sgm_0 \eql 
 =-i\kp\xi_1\gm_{(3)}^m\zeta_0\der_m\sgm_0, \nonumber\\
 \dlt_{\xi_1}^{Q_1}\phi \eql 
 =-i\kp\xi_1\gm_{(3)}^m\zeta_0\der_m\phi. 
 \label{dltQ1_st}
\eea

For the broken Lorentz symmetry, 
\bea
 \dlt_v^K\rho_0 \eql -\kp^{-1}v^mx_m+\cO(\kp), \nonumber\\
 \dlt_v^K\zeta_{0\alpha} \eql -\frac{i}{2}\kp^{-1}v^m
 \brkt{\gm_{(3)m}\tht_2}_\alpha+\cO(\kp), \nonumber\\
 \dlt_v^K\lmd_0^m \eql \kp^{-1}v^m+\cO(\kp), \nonumber\\
 \dlt_v^K\sgm_0 \eql \cO(\kp), \nonumber\\
 \dlt_v^K\phi \eql \kp v^m\brc{\brkt{-\rho_0+\frac{1}{2}
 \zeta_0\gm_{(3)}^n\gm_{(3)m}\tht_2}\der_n\phi
 +\frac{i}{2}\zeta_0\gm_{(3)m}D_2\phi}. 
 \label{dltK_st}
\eea

For the $U(1)_R$ symmetry, 
\bea
 \dlt_r^R\rho_0 \eql r\brc{\kp^{-1}\tht_2^2-\kp\zeta_0^2
 -i\kp\zeta_0\gm_{(3)}^m\tht_2\der_m\rho_0-\kp\zeta_0D_2
 \rho_0}, \nonumber\\
 \dlt_r^R\zeta_{0\alpha} \eql r\brc{-\kp^{-1}\tht_{2\alpha}
 -i\kp\zeta_0\gm_{(3)}^m\tht_2\der_m\zeta_{0\alpha}
 -\kp\zeta_0D_2\zeta_{0\alpha}}, \nonumber\\
 \dlt_r^R\lmd_0^m \eql r\brc{-i\kp\zeta_0\gm_{(3)}^n\tht_2
 \der_n\lmd_0^m-\kp\zeta_0D_2\lmd_0^m}, \nonumber\\
 \dlt_r^R\sgm_0 \eql r\brc{\kp^{-1}-i\kp\zeta_0\gm_{(3)}^m
 \tht_2\der_m\sgm_0-\kp\zeta_0D_2\sgm_0}, \nonumber\\
 \dlt_r^R\phi \eql r\brc{-i\kp\zeta_0\gm_{(3)}^m\tht_2
 \der_m\phi-\kp\zeta_0D_2\phi}. \label{dltR_st}
\eea

Here, $a$, $\xi_{1\alpha}$, $v^m$ and $r$ are transformation 
parameters, and $\phi$ denotes a matter field. 

\subsection{Inverse Higgs effect}
The NG superfields introduced in Eq.(\ref{hatOmega}) are not 
independent. 
The relation between them are obtained by the following 
covariant constraints~\cite{ivsHiggs,manohar}. 
\be
 \hat{\cD}_{2\alpha}\rho_0=0, \;\;\;
 \hat{\cD}_{2\alpha}\zeta_0^\beta=0, \;\;\;
 \hat{\cD}_m\rho_0=0. 
\ee
Here $\hat{\cD}_m$ and $\hat{\cD}_{2\alpha}$ are covariant 
derivatives in the presence of the NG superfields, 
which are derived from 
the Cartan one-form~$\hat{\Omega}^{-1}{\rm d}\hat{\Omega}$. 

Solving these constraints, we can express all NG superfields 
in terms of a single superfield~$\rho_0$. 
\bea
 \zeta_{0\alpha} \eql -\frac{1}{2}D_{2\alpha}\rho_0+\cO(\kp^2), 
 \nonumber\\
 \lmd_0^m \eql iD_2\gm_{(3)}^m\zeta_0+\cO(\kp^2)
 =-\der^m\rho_0+\cO(\kp^2), \nonumber\\
 \sgm_0 \eql \frac{1}{2}D_2^\alpha\zeta_{0\alpha}+\cO(\kp^2) 
 =-\frac{1}{4}D_2^2\rho_0+\cO(\kp^2). 
\eea

\end{document}